\documentclass[10pt,letterpaper]{article}
\usepackage[top=0.85in,left=2.75in,footskip=0.75in]{geometry}

\usepackage{amsmath,amssymb}

\usepackage{changepage}

\usepackage[utf8x]{inputenc}

\usepackage{textcomp,marvosym}

\usepackage{cite}

\usepackage{nameref,hyperref}

\usepackage[right]{lineno}

\usepackage{microtype}
\DisableLigatures[f]{encoding = *, family = * }

\usepackage[table]{xcolor}

\usepackage{array}

\newcolumntype{+}{!{\vrule width 2pt}}

\newlength\savedwidth

\raggedright
\usepackage[aboveskip=1pt,labelfont=bf,labelsep=period,justification=raggedright,singlelinecheck=off]{caption}

\makeatletter
\renewcommand{\@biblabel}[1]{\quad#1.}
\makeatother

\usepackage{lastpage,fancyhdr,graphicx}
\usepackage{epstopdf}
\fancyheadoffset[L]{2.25in}
\fancyfootoffset[L]{2.25in}
\usepackage{xspace}
\usepackage[squaren]{SIunits}

\newcommand{\me}[1]{\mathrm{e}^{#1}}
\newcommand{\ebov}{EBOV\xspace}
\newcommand{\dift}[1]{\frac{\mathrm{d}#1}{\mathrm{d}t}}
\newcommand{\Vinf}{\ensuremath{V_\text{inf}}\xspace}
\newcommand{\pinf}{\ensuremath{p_\text{inf}}\xspace}
\newcommand{\cinf}{\ensuremath{c_\text{inf}}\xspace}
\newcommand{\Vtot}{\ensuremath{V_\text{tot}}\xspace}
\newcommand{\ptot}{\ensuremath{p_\text{tot}}\xspace}
\newcommand{\ctot}{\ensuremath{c_\text{tot}}\xspace}
\newcommand{\tcid}{\ensuremath{\text{TCID}_{50}}\xspace}
\newcommand{\qpcr}{RT-qPCR\xspace}
\newcommand{\func}{\ensuremath{\mathcal{F}}}

\begin{document}
\vspace*{0.2in}

\begin{flushleft}
{\Large
\textbf\newline{Quantification of Ebola virus replication kinetics in vitro} 
}
\newline
\\
Laura E.\ Liao\textsuperscript{a,1}
Jonathan Carruthers\textsuperscript{b,1}
Sophie J.\ Smither\textsuperscript{c}
CL4 Virology Team\textsuperscript{c,2}
Simon A. Weller\textsuperscript{c}
Diane Williamson\textsuperscript{c}
Thomas R. \ Laws\textsuperscript{c}
Isabel Garc\'{i}a-Dorival\textsuperscript{d}
Julian Hiscox\textsuperscript{d}
Benjamin P.\ Holder\textsuperscript{e}
Catherine A.\ A.\ Beauchemin\textsuperscript{f,g}
Alan S.\ Perelson\textsuperscript{a}
Mart\'{i}n L\'{o}pez-Garc\'{i}a\textsuperscript{b}
Grant Lythe\textsuperscript{b}
John Barr\textsuperscript{h}
Carmen Molina-Par\'{i}s\textsuperscript{b,3}
\\

\bigskip
\textbf{a} Theoretical Biology and Biophysics, Los Alamos National Laboratory, Los Alamos, NM, USA 87545
\\
\textbf{b} Department of Applied Mathematics, School of Mathematics, University of Leeds, Leeds LS2 9JT, UK
\\
\textbf{c} Defence Science and Technology Laboratory, Salisbury SP4 0JQ, UK
\\
\textbf{d} Institute of Infection and Global Health, University of Liverpool, Liverpool, L69 7BE, UK
\\
\textbf{e} Department of Physics, Grand Valley State University, Allendale, MI, USA 49401
\\
\textbf{f} Department of Physics, Ryerson University, Toronto, ON, Canada M5B 2K3
\\
\textbf{g} Interdisciplinary Theoretical and Mathematical Sciences (iTHEMS) Research Program at RIKEN, Wako, Saitama, Japan, 351-0198
\\
\textbf{h} School of Molecular and Cellular Biology, University of Leeds, Leeds LS2 9JT, UK
\\
\bigskip

%
%
1 These authors contributed equally to this work.




2 Membership list can be found in the Acknowledgments section.

3 carmen@maths.leeds.ac.uk

\end{flushleft}
\section*{Abstract}
Mathematical modelling has successfully been used to provide quantitative descriptions of many viral infections, but for the Ebola virus, which requires biosafety level 4 facilities for experimentation, modelling can play a crucial role. Ebola modelling efforts have primarily focused on \textit{in vivo} virus kinetics, e.g., in animal models, to aid the development of antivirals and vaccines. But, thus far, these studies have not yielded a detailed specification of the infection cycle, which could provide a foundational description of the virus kinetics and thus a deeper understanding of their clinical manifestation. Here, we obtain a diverse experimental data set of the Ebola infection \textit{in vitro}, and then make use of Bayesian inference methods to fully identify parameters in a mathematical model of the infection. Our results provide insights into the distribution of time an infected cell spends in the eclipse phase (the period between infection and the start of virus production), as well as the rate at which infectious virions lose infectivity. We suggest how these results can be used in future models to describe co-infection with defective interfering particles, which are an emerging alternative therapeutic.

\section*{Author summary}
The two deadliest Ebola epidemics have both occurred in the past five years, with one of these epidemics still ongoing. Mathematical modelling has already provided insights into the spread of disease at the population level as well as the effect of antiviral therapy in Ebola-infected animals. However, a quantitative description of the replication cycle is still missing. Here, we report results from a set of \textit{in vitro} experiments involving infection with the Ecran strain of Ebola virus. By parameterizing a mathematical model, we are able to determine robust estimates for the duration of the replication cycle, the infectious burst size, and the viral clearance rate.

\section*{Introduction}
The world's second largest Ebola outbreak is currently underway in the Democratic Republic of Congo. Ebola virus (\ebov) causes severe and fatal disease with death rates of up to 90\%~\cite{feldmann11}. There is an urgent need to prevent and treat \ebov infections, but no antiviral drugs or monoclonal antibodies have been approved in Africa, the EU, or the US. Recently the first \ebov vaccine has been approved by European regulators~\cite{callaway19}. Experimental therapies~\cite{rojek17}, including antiviral drugs (remdesivir~\cite{dornemann17} and favipiravir~\cite{sissoko16, carazoperez17}) and a cocktail of monoclonal antibodies (ZMapp)~\cite{davey16}, have been assessed in the 2013--2016 West Africa Ebola virus disease outbreak. Other promising monoclonal antibody therapies, called mAb114 and REGN-EB3, have been deployed in the current 2018--2019 Kivu Ebola epidemic~\cite{maxmen18}. A better understanding of the precise infection kinetics of \ebov is warranted.

Mathematical modelling of viral dynamics has provided a quantitative understanding of within-host viral infections, such as HIV~\cite{ho96}, influenza~\cite{baccam06}, Zika~\cite{best17}, and more recently, \ebov. Mathematical modelling studies have analyzed the plasma viral load dynamics of \ebov-infected animals (mice~\cite{madelain15}, non-human primates~\cite{guedj18,madelain18}) while under therapy with favipiravir, and have identified estimates of favipiravir efficacy and target drug concentrations. In addition, mechanistic models of innate and adaptive immune responses were used to provide an explanation of \ebov infection dynamics in non-human primates~\cite{madelain18}, and of differences between fatal and non-fatal cases of human infection~\cite{martyushev16}. Moreover, mathematical models have been used to predict the effect of treatment initiation time on indicators of disease severity~\cite{madelain15,martyushev16} and survival rates~\cite{madelain18}, to predict the clearance of EBOV from seminal fluid of survivors~\cite{sissoko17}, and to theoretically explore treatment of \ebov-infected humans with antivirals that possess different mechanisms of action (i.e., nucleoside analog, siRNA, antibody)~\cite{martyushev16}.

Alongside the progress made in understanding within-host infections, a complementary view of infection can be provided by mathematical modelling of infections at the \textit{in vitro} level. Combined with \textit{in vitro} time course data, mathematical models (MMs) have provided a detailed quantitative description of the viral replication cycle of influenza A virus~\cite{beauchemin18coisb,mohler05}, SHIV~\cite{iwami12,beauchemin17}, HIV~\cite{iwami15}, and other viruses~\cite{beauchemin19, gonzalezparra18,fukuhara13,gonzalezparra18rota,wethington18}. Such studies yield estimates of key quantities such as the basic reproductive number (defined as the number of secondary infections caused by one infected cell in a population of fully susceptible cells), half-life of infected cells, and viral burst size, which cannot be obtained directly from data~\cite{iwami12front}. In the context of \textit{in vitro} infections, parameterized MMs have been used to predict the outcome of competition experiments between virus strains~\cite{holder12compete, song12, paradis15} (i.e., which strain dominates in a mixed infection), map differences in genotype to changes in phenotype~\cite{holder12compete, paradis15} (e.g., associate a single mutation to ten-fold faster viral production), quantify fitness differences between virus strains~\cite{simon16, iwanami17} (e.g., which strain has a larger infectious burst size), quantify the contribution of different modes of transmission (cell-to-cell versus cell-free)~\cite{iwami15}, and identify the target of antiviral candidates~\cite{ikeda15} (e.g., whether a drug inhibits viral entry or viral production). One prior study~\cite{nguyen15} utilized \textit{in vitro} infection data from the literature to estimate \ebov infection parameters, but had several parameter identifiability issues due to insufficient data.

Our goal is to obtain robust estimates of viral infection parameters that characterize the \ebov replication cycle. We follow a mathematical modelling approach that has been successfully applied in the analysis of other viral infections \textit{in vitro}~\cite{beauchemin18coisb}. To this end, we performed a suite of \textit{in vitro} infection assays (single-cycle, multiple-cycle, and viral infectivity decay assays) using \ebov and Vero cells, and collected detailed extracellular infectious and total virus time courses. The viral kinetic data were simulated with a multicompartment ordinary differential equation MM, and posterior distributions of the MM parameters were estimated using a Markov chain Monte Carlo (MCMC) approach. We estimate that one EBOV-infected cell spends $\sim\unit{30}{\hour}$ in an eclipse phase before it releases infectious virions at a rate of \unit{13}{\per\hour}, over its infectious lifetime of $\sim\unit{83}{\hour}$. The number of infectious virions produced over an infected cell's lifetime is $\sim1000$, with an estimated basic reproductive number of $\sim600$. We also discuss challenges in collecting other types of virus dynamic data (e.g., intracellular viral RNA or cell counts).

\section*{Results}
\subsubsection*{Ebola Virus Kinetics In Vitro}

\begin{figure*}[t!]
\centering
\includegraphics[width=\linewidth]{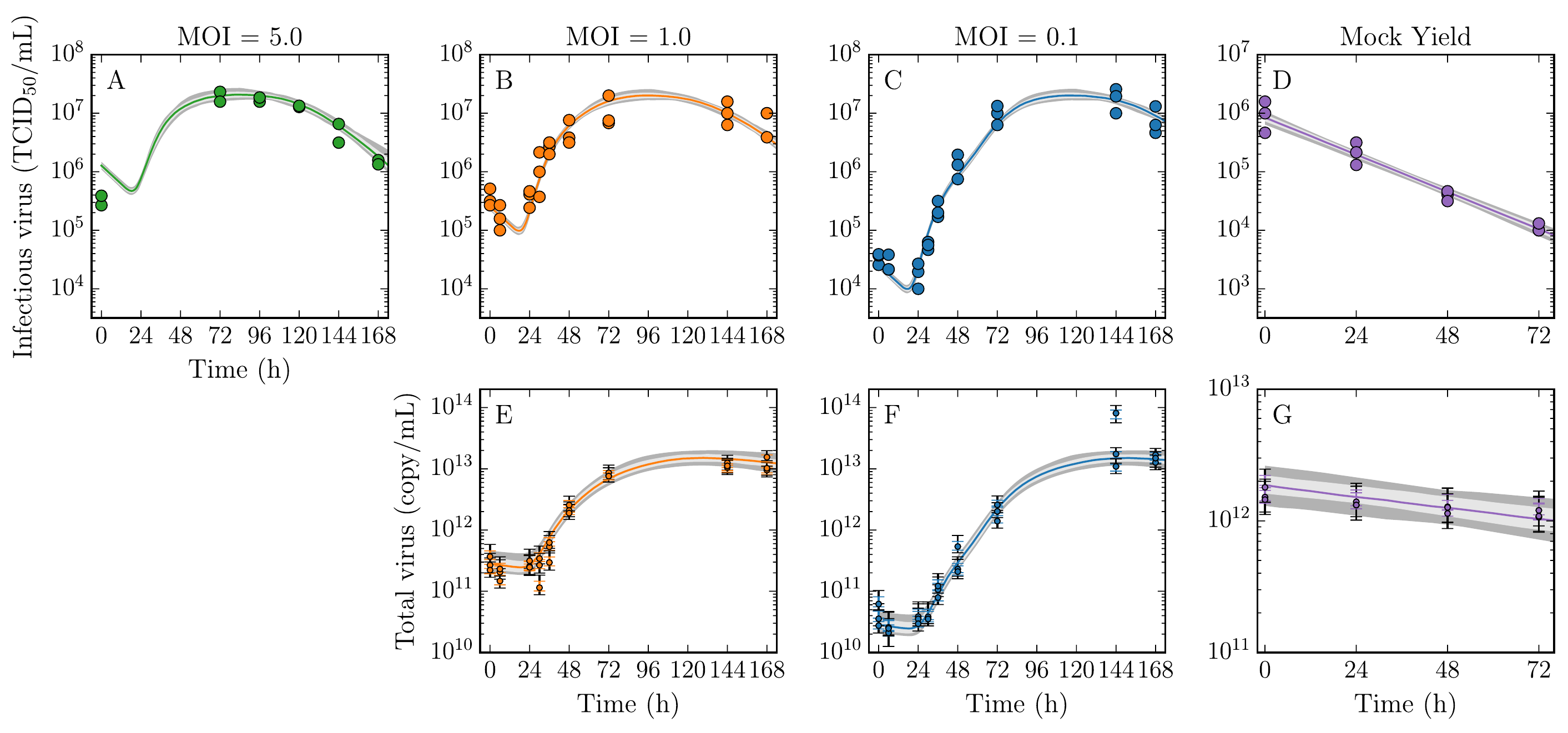}
\caption{\textbf{Kinetics of \ebov infection \textit{\textbf{in vitro}} and mock yield assays.} %
Vero cell monolayers were infected with \ebov at a multiplicity of infection (MOI) 5, 1, or \unit{0.1}{\tcid \per cell}, as indicated. At various times post-infection, the infectious (\tcid/mL; A--C) and total virus (copy/mL; E--G) in the supernatant were determined. A mock yield assay was also performed to quantify the decay of infectious (D) and total virus (G). In each assay, the experimental data (circles) were collected either in duplicate (MOI 5) or triplicate (all other assays). Note that the total virus concentration collected in the MOI 5 infection was omitted from the analysis due to inconsistencies in the peak value (S1 Appendix, Fig.~A). 
The lines represent the pointwise median of the time courses simulated from our MM, which are bracketed by 68\% (light grey) and 95\% (dark grey) credible regions (CR). These data were used to extract the posterior probability likelihood distributions of the infection parameters (Fig.~\ref{fig:PLDs}). Note that parameters of the calibration curve used to convert cycle threshold values (Ct) to total virus (copy/mL) were also estimated (Fig.~\ref{fig:stdcurve}). The variability introduced from this conversion is shown by two error bars on each total virus data point, indicating the 68\% (same colour) and 95\% (black) CR. %
} 
\label{fig:fits}
\end{figure*}

Vero cell monolayers were infected with \ebov at a multiplicity of infection (MOI) of 5, 1, \unit{0.1}{\tcid\per cell}. Infectious and total virus concentrations were determined from extracellular virus harvested from the supernatant of each well at various times post-infection (Fig.~\ref{fig:fits} A--C, E---F). At the start of infection, the virus concentrations do not rise for some time, reflecting the time it takes for viral entry, replication and release. After \unit{24}{\hour}, the virus concentrations grow exponentially as infected cells begin producing virus. When all cells in the well are infected, the virus concentrations peak at approximately \unit{2\times10^7}{\tcid\per\milli\liter} and \unit{10^{13}}{copy\per\milli\liter} and the peak is sustained for $\sim\unit{72}{\hour}$. Thereafter, the virus concentrations decline when virus production ceases, presumably due to the death of infected cells. Additionally, the kinetics of viral infectivity decay and virus degradation were assessed with a mock yield assay (Fig.~\ref{fig:fits} D, G). In the mock yield assay, an inoculum of virus was incubated in wells under the same conditions as the growth assays, but in the absence of cells, and sampled over time.
 
\subsubsection*{Mathematical Model of Viral Infection and Parameter Estimates}

The \textit{in vitro} \ebov infection kinetics were captured with a MM that has been used successfully in past works to capture influenza A virus infection kinetics \textit{in vitro} \cite{paradis15, holder12compete, simon16}. The MM is given by the system of ordinary differential equations:
\begin{align}
\dift{T} &= -\beta T \Vinf \nonumber\\
\dift{E_1} &= \beta T \Vinf -\frac{n_E}{\tau_E} E_1 \nonumber\\
\dift{E_{i=2,3,...,n_E}} &= \frac{n_E}{\tau_E} \left( E_{i-1} - E_i \right) \nonumber \\
\dift{I_1} &= \frac{n_E}{\tau_E} E_{n_E} - \frac{n_I}{\tau_I} I_1 \label{eqnMM}\\
\dift{I_{j=2,3,...,n_I}} &= \frac{n_I}{\tau_I} \left( I_{j-1} - I_j \right) \nonumber\\
\dift{\Vinf} &= \pinf \sum_{j=1}^{n_I} I_j - \cinf \Vinf \nonumber\\
\dift{\Vtot} &= \ptot \sum_{j=1}^{n_I} I_j - \ctot \Vtot \nonumber
\end{align}
In this MM, susceptible uninfected target cells $T$ can be infected by infectious virus \Vinf with infection rate constant $\beta$, and subsequently enter the non-productive eclipse phase $E_{i=1,\ldots,n_E}$, followed by a transition into the productively infectious phase $I_{j=1,\ldots,n_I}$. The eclipse and infectious phases are divided into a number of compartments given by $n_E$ and $n_I$, respectively, such that the time spent in each phase follows an Erlang distribution with an average duration of $\tau_{E,I} \pm \frac{\tau_{E,I}}{\sqrt{n_{E,I}}}$. While cells are in the infectious phase, they produce infectious (total) virus \Vinf (\Vtot) at a rate \pinf (\ptot), which lose infectivity (viability) at rate \cinf (\ctot). The MM Eq.~\eqref{eqnMM} captures both infectious virus, quantified by \tcid measurements of supernatant samples, and total virus, quantified by quantitative, real-time, reverse transcriptase PCR (hereafter, \qpcr). The latter experimental quantity was obtained by converting cycle threshold (Ct) values from \qpcr to copy number (Fig.~\ref{fig:stdcurve}) using Eq.~\eqref{eqnPCRsample} (Methods).

\begin{figure*}[t]
\centering
\includegraphics[width=\linewidth]{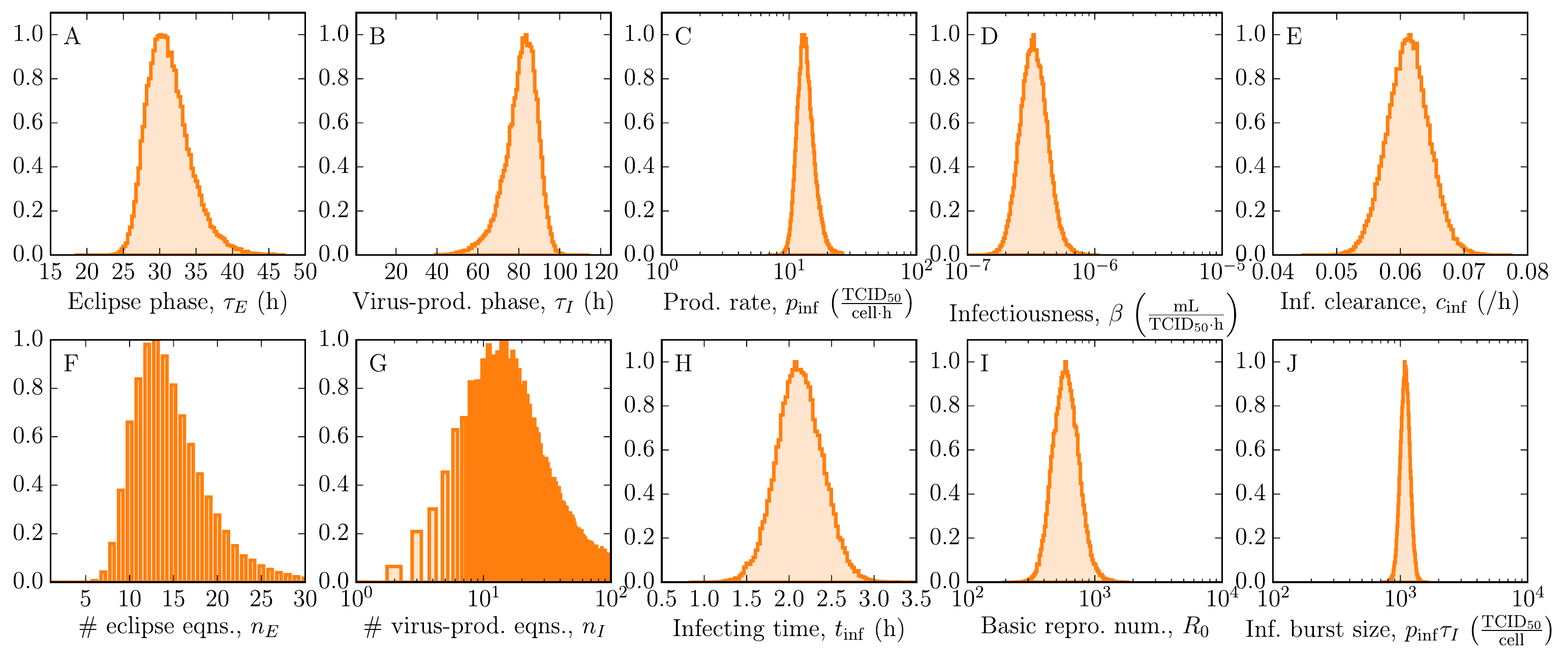}
\caption{\textbf{Estimated parameter distributions of \ebov infection \textit{\textbf{in vitro}}.} %
Posterior probability likelihood distributions (PostPLDs) of parameters in the MM (A--G) were estimated using MCMC and the data in Fig.~\ref{fig:fits}. Secondary parameters were derived from these estimates (H--J). Note that the PostPLDs corresponding to the number of eclipse and infectious phase compartments are integer-valued. The remaining PostPLDs of parameters describing the total virus and calibration curve are in S1 Appendix, Fig.~B. 
} 
\label{fig:PLDs}
\end{figure*}

Predicted virus time courses from the MM are shown in Fig.~\ref{fig:fits} where the solid lines represent the pointwise median and the grey bands show narrow 95\% credible regions (CR), indicating that the MM reproduces the viral kinetic data well. Using a Markov chain Monte Carlo (MCMC) approach, we obtained posterior probability likelihood distributions (PostPLDs) for each of the MM parameters (Fig.~\ref{fig:PLDs}). Narrow PostPLDs were extracted with mild correlations between parameters (S1 Appendix, Fig.~C), 
indicating good practical identification of all parameters.

\subsubsection*{A Quantitative Description of the EBOV Lifecycle}
 
The MCMC analysis gives us the following quantitative description of the \ebov lifecycle within Vero cells. An \ebov-infected Vero cell spends approximately \unit{30}{\hour} with a 95\% credible region of [\unit{26}{\hour}, \unit{37}{\hour}] in the eclipse phase before progeny \ebov successfully bud. Subsequently, infectious virus is produced at a rate of 13 [10, 20] virions per cell per hour over a duration of \unit{83}{\hour} [\unit{64}{\hour}, \unit{95}{\hour}] before virus production ceases due to cell death. An infectious burst size of 1096 progeny virions [1000, 1259] is released from each infected cell over its virus-producing lifetime. Once infectious virus enters the cell culture medium, they lose infectivity at a rate of \unit{0.06}{\per\hour} [\unit{0.055}{\per\hour}, \unit{0.068}{\per\hour}], which is comparable to other viruses such as influenza A virus~\cite{simon16} or SHIV~\cite{iwami12}. Overall, the \textit{in vitro} spread of infection is rapid, as characterized by an infecting time of \unit{2}{\hour} [\unit{1.6}{\hour}, \unit{2.7}{\hour}], which is defined as the time required for a single virus-producing cell to infect one more~\cite{holder11delay}. These dynamics imply a large basic reproductive number of \unit{589} [398, 1000], which is defined as the number of secondary infections caused by a single infected cell in a population of fully susceptible cells.

Notably, we find that the durations of both the eclipse and infectious phases follow a normal-like distribution, as given by $n_E$ of 13 [8, 23] and $n_I$ of 14 [3, 85]. This implies that the eclipse phase comprises a sequence of many distinct steps of short duration, without any one step lasting significantly longer than the rest. Likewise, the same interpretation applies to the infectious phase. The normal-like distribution of the eclipse phase resembles that of influenza A virus~\cite{paradis15}, but contrasts with the fat-tailed eclipse phase distribution of SHIV~\cite{beauchemin17} which is likely due to a process in the phase that is longer than the rest (e.g., integration). Moreover, neither the eclipse nor infectious phase are exponentially distributed ($n=1$) as is commonly assumed in analyses with MMs. Such an assumption has been shown to impact estimates of antiviral efficacy that are based on patterns of viral load decay under simulated therapy in HIV patients~\cite{beauchemin17}.

\section*{Discussion}
In this work we performed time-course Ebola virus (EBOV) infection experiments at multiple MOIs \textit{in vitro} and applied MCMC methods to precisely parameterize a mathematical model (MM) of the infection. We extracted fundamental quantities concerning the timing and viral production of \ebov replication. This theoretical-experimental approach maximized the output of the costly and difficult experiments, which must be performed in biosafety level 4 facilities. Previous studies of the \ebov lifecycle rely on safer virus-like particles~\cite{biedenkopf2017modeling}. The only previously known MM of \ebov infection \textit{in vitro}~\cite{nguyen15} is restricted in its use due to problems with parameter identifiability; specifically, the existence of strong correlations between parameters, such as the rates of virus degradation and virus production. By obtaining a more complete set of experimental observations, we have provided the first detailed quantitative characterization of \ebov infection kinetics. %

Some of our estimates of timescales in the EBOV infection kinetics fill gaps in the knowledge of this virus, while others expose some tension with prior mathematical modelling work. The eclipse phase, excluded from the previous \textit{in vitro} MM~\cite{nguyen15}, has been found to be a significant part of the replication cycle. Lasting approximately \unit{30}{\hour}, it is longer than the eclipse phase for influenza A virus and HIV infections in humans (4--\unit{24}{\hour})~\cite{baccam06,dixit2004estimates}. Although the eclipse phase is included in existing MMs of \ebov-infected animals, its duration has never been estimated, and the assumed values used in these studies were considerably shorter than the value we identify here~\cite{madelain15,madelain18}. Moreover, the observation that the length of the eclipse phase follows an Erlang distribution is contrary to these previous MMs, where it has been represented more simply as an exponentially distributed time. These MMs also fix the value of the decay rate of infectious virus to ensure that other parameters remain identifiable~\cite{madelain15}. Here, the robust estimate of this decay rate demonstrates the benefit of performing a mock yield assay. Existing MMs of \textit{in vivo} \ebov infection in humans and non-human primates provide considerably shorter estimates of the infection cycle (12.5--\unit{15.3}{\hour}) compared to the estimate of \unit{114}{\hour} ($\tau_E + \tau_I$) obtained here~\cite{madelain15,martyushev16}. Such a difference is likely attributed to the inclusion of an implicit immune response in these \textit{in vivo} models, thereby accounting for the enhanced clearance of infected cells by immune cells, such as CD8$^+$ T~cells \cite{gupta2005cd8}. This also explains why a faster viral decay rate can be expected \textit{in vivo}, and subsequently why estimates of the basic reproduction number are greater here than those obtained from \textit{in vivo} MMs (5.96-9.01)~\cite{madelain15,martyushev16}. It remains to be determined whether Vero cells are representative of the cells targeted by \ebov \textit{in vivo}, but by understanding EBOV replication in Vero cells, we have a foundation from which more complex cell culture models might be developed. 

In addition to virus measurements, previous studies have included susceptible and infected cell measurements to fully parameterize the MM and obtain robust estimates of the viral kinetics parameters~\cite{iwami12, schulzehorsel09}. We initially set out to obtain a more diverse data set that also included the kinetics of dead cells and intracellular RNA over the course of infection, but encountered unexpected challenges. To quantify cell viability, we treated infected monolayers at various times with Trypan blue, which stains cells that have lost the ability to exclude dye. Unfortunately, we were unable to associate this marker of cell death to a stage of the viral lifecycle in our MM without making additional assumptions. Ultimately, when we extended the MM to include these data, the newly introduced parameters were dependent on these assumptions, and the extracted values of the original parameters were largely unaffected (S1 Appendix). To determine intracellular viral kinetics, the supernatants from infected cell cultures were removed and the remaining monolayers were washed and trypsinized for quantification via \tcid assay and \qpcr. These samples showed a high level of EBOV RNA and \tcid as early as 4 hours post-infection, which remained at a constant level up to 1 day post-infection, but rose thereafter (S1 Appendix). Additionally, the ratio of RNA-to-\tcid resembled the ratio observed in the supernatant. Thus, these measurements likely reflect the large amount of cell-associated virions that remained after washing, effectively obscuring the intracellular RNA signal. 

While a highly-controlled \textit{in vitro} system was necessary to achieve our precise characterization of the EBOV infection kinetics, the applicability of these results to a clinical situation is not immediately obvious, and represents a serious limitation of the study. Nevertheless, our findings have some relevance to understanding the EBOV infection \textit{in vivo}. \ebov initially replicates within macrophages and dendritic cells in subcutaneous and submucosal compartments, but dissemination in the blood results in the infection of multiple organs throughout the body~\cite{chertow2020modeling}. Many different cell types are infected with varying susceptibility to infection, as well as varying levels of viral replication. While the infection unfolds, EBOV blocks IFN production early on~\cite{carazoperez17,edwards2016differential}. In this sense, studying the infection of Vero cells---which are IFN-deficient---narrowly models the infection of one type of epithelial cell during the early stages of an EBOV infection \textit{in vivo}. 

Vero cells serve as a standard host cell for replication and are widely used for testing antivirals \textit{in vitro}~\cite{postnikova18}, as well as in the development of viral vaccines~\cite{barrett09,barrett2017vero,paillet2009suspension}. Mathematical modelling of EBOV infections \textit{in vitro} using Vero cells has relevance to such applications, particularly in the study of emerging therapeutics. While we provided a quantitative depiction of extracellular infection by EBOV as a valuable first step, we envisioned that the MM could be extended to include intracellular viral RNA kinetics had the appropriate data been collected. Such multiscale modelling approaches have been used to provide insight into virus growth and also to the understanding of direct-acting antivirals~\cite{guedj13,heldt13,quintela18,zitzmann18}. We hope that these experiences might help guide future efforts to obtain informative cell and intracellular data. 

As an alternative antiviral strategy, there has been renewed interest in pursuing defective interfering particles (DIPs)~\cite{rezelj18} of highly pathogenic viruses. A DIP is a viral particle that contains defective interfering RNA (DI RNA), which can be a shortened version of the parent genome that renders a DIP replication-incompetent on its own (because it may lack the gene for an essential viral component such as viral polymerase), but also elicits virus-interfering properties. Within a cell co-infected by both DIPs and virus, the DI RNA has a replicative advantage over the full-length RNA and outcompetes it to produce more DIPs than virus progeny, effectively reducing the infectious virus yield. EBOV DI RNA has been observed~\cite{calain99} but much remains to be understood. Like with any other antiviral, MMs can be used to determine the efficacy and mechanism of action of candidate DI RNAs, and to explore the impact of dose and timing~\cite{martyushev16}. In particular, our estimates of EBOV infection kinetics parameters are directly applicable to future mathematical modelling of the interactions between EBOV and EBOV DIPs \textit{in vitro}. Our estimated EBOV infection parameters may also describe certain aspects of EBOV DIP infection. For example, since DIPs have the same viral proteins and capsid as virions, they would infect cells with the same infection rate constant, $\beta$. Since DIPs also piggyback on the virus' replication cycle, we might expect the same eclipse and infectious phase lengths ($\tau_E, \tau_I$) in a DIP and virus co-infected cell. 

In summary, the MM described here characterizes the replication cycle of EBOV in a quantitative manner that will be beneficial for those creating \textit{in vitro} models to aid the development of antivirals and vaccines. We have made use of a valuable set of \textit{in vitro} results, carefully considering the structure of the MM in order to maximize the information we can extract from them.

\section*{Materials and Methods}
\subsection*{Cells and Virus}
Vero C1008 cells (ECACC Cat. No.85020206) were obtained from Culture Collection, Public Health England, UK. Vero C1008 cells were maintained in Dulbecco's minimum essential media supplemented with 10\% (v/v) foetal calf serum, 1\% (v/v) L-glutamine and 1\% (v/v) penicillin/streptomycin (Sigma). For experimental purposes, the foetal calf serum concentration was reduced to 2\% (v/v).

Ebola virus \textit{H. sapiens}-tc/COD/1976/Yambuku-Ecran, hereafter referred to as \ebov was used in all studies. This virus, previously known as EBOV ``E718'' \cite{kuhn14} was supplied by Public Health England. Passage 5 material was used to infect Vero C1008 cells. Virus was harvested on day 5 post-inoculation and titrated to produce a working stock at \unit{10^7}{\tcid\per\milli\liter}. 

\subsection*{Quantification of Virus} 
\ebov was titrated in 96-well plates using the endpoint fifty percent tissue culture infectious dose (\tcid) assay \cite{smither13}. Briefly, virus was ten-fold serially diluted in 96 well plates of Vero C1008 cells. After one week of incubation at \unit{37}{\degree\Celsius}/5\% $\mathrm{CO}_2$, all wells were observed under the microscope and scored for presence or absence of cytopathic effects. The 50\% endpoint was then calculated using the method of Reed \& Muench \cite{reed38}. RNA extractions were performed using the QiAMP Viral RNA Mini Kit (Qiagen, UK). Two \unit{50}{\micro\liter} elutions were performed for each sample to increase the volume available for RT-PCR. 

The genetic material of \ebov was quantified using the RealStar\textregistered\ Filovirus Screen RT-PCR Kit (Altona diagnostics, Country) following the instructions of the manufacturer. This assay has been performed many times against a standard curve of plasmid containing the L gene from \ebov. The number of genomes can be estimated from the Ct values as described in Eq.~\eqref{eqnPCRsample}. In this context, the number of genomes might consist of incomplete negative sense RNA molecules encoding this sequence of the L gene. However, we do not believe that these will be common ($<5\%$) based upon observations made with next generation sequencing (paper in preparation). MOI 5 experiments were analysed using a BIORAD CFX Connect – Real Time System, while samples for the remaining MOIs were analysed using a QuantStudio 7 Flex Real-Time PCR System. Signal from control RNA was compared between experiments and machines and we found no evidence of differences. The parameters of the calibration curve required to convert Ct values to total virus used samples from the MOI 5 experiments.

\subsection*{Infections}
Twenty-four-well plates were seeded with Vero C1008 cells at \unit{10^5}{cells\per\milli\liter}. \ebov was added at MOIs of either 5, 1, or 0.1. Vero cells were grown to 90\% confluence for all infections. The cell culture medium was not changed during the experiment and all cultures reached confluence within \unit{24}{\hour} (S1 Appendix). At pre-determined intervals post-infection samples were taken by aspiration of supernatant from wells. Samples were stored at \unit{-80}{\degree\Celsius} prior to enumeration by TCID50 assay and RNA extraction for PCR. Note that the RNA from the MOI 5 infection was omitted from further analysis due to inconsistencies in the peak viral RNA, compared to the MOI 1 and 0.1 infections (S1 Appendix, Fig.~A). 
The viability of Vero cells in the absence of infection is not known under these conditions, however, we have observed these cells for 168 h at 24 h intervals and observed only occasional cells that can be stained with the viability stain Trypan blue.

\subsection*{Mock yield or infectivity decay assay}
EBOV was added to twenty-four-well plates at a final estimated density of $5 \times 10^5$ $\tcid$. At pre-determined intervals post-infection samples were taken by aspiration of supernatant from wells. Samples were stored at \unit{-80}{\degree\Celsius} prior to enumeration by $\tcid$ assay and RNA extraction for PCR.

\subsection*{Construction of the standard \qpcr curve}

\begin{figure*}[t!]
\centering
\includegraphics[width=0.3\linewidth]{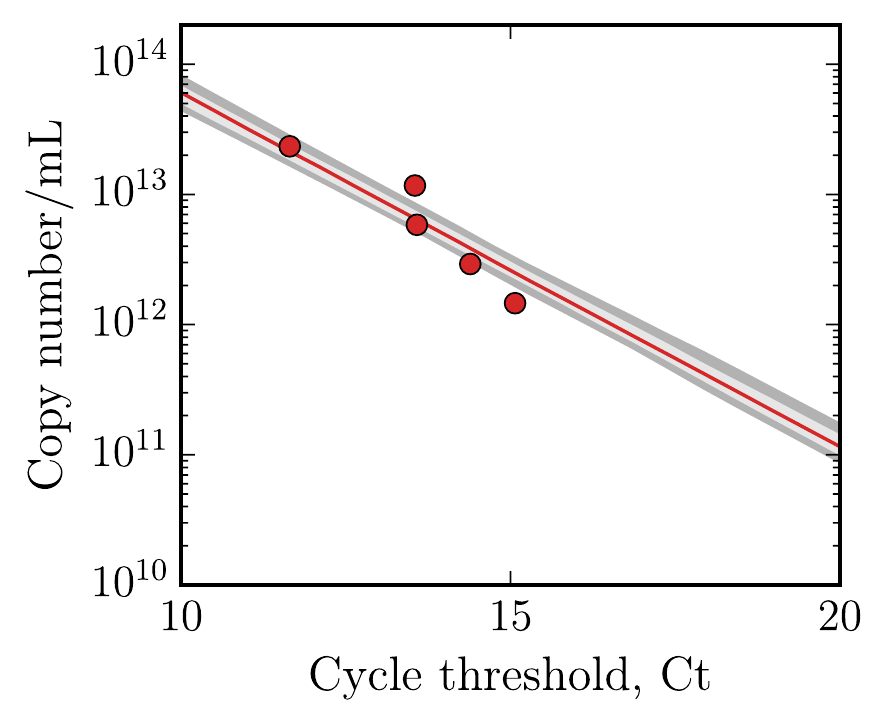}
\caption{\textbf{Standard \qpcr curve.} %
Cycle threshold values (Ct) were converted to total virus (copy/mL) using the above calibration curve, where the parameters of the curve were estimated as a part of the analysis. The lines represent the pointwise median of the time courses simulated from our MM, which are bracketed by 68\% (light grey) and 95\% (dark grey) CR. %
} 
\label{fig:stdcurve}
\end{figure*}

The concentration of viral genome copies (copy/mL) in a standard sample $i$ ($V_{\text{STD},i}$) and the number of doubling \qpcr cycles ($C_{t,\text{STD},i}$) required for this concentration of copies to reach an arbitrarily fixed, chosen threshold concentration ($Q_t$), are linked by the equation
\begin{align}
Q_t &= V_{\text{STD},i}\ (2\varepsilon)^{C_{t,\text{STD},i}} \nonumber\\
\ln(V_{\text{STD},i}) &= \underbrace{\ln(Q_t)}_{y\text{-intercept}} - \underbrace{\ln(2\varepsilon)}_\text{slope}\ C_{t,\text{STD},i} \label{eqnPCRstd}
\end{align}
where $\varepsilon$ is the efficacy of the \qpcr doubling, which should ideally be equal to one (i.e., exactly doubles at each cycle) but can vary about this value. In constructing the standard curve, we took five standard samples ($V_{\text{STD},i=1...5}$) with known copy concentrations (via their mass) and determined their corresponding $C_{t,\text{STD},i}$. These data are shown in Fig.~\ref{fig:stdcurve}. 
 
\subsection*{Conversion of sample \qpcr $C_t$ values into $\Vtot$}
In quantifying the concentration of total virus, \Vtot (copy/mL), in the extracellular virus samples collected from infection experiments, Eq.~\eqref{eqnPCRstd} was used as follows
\begin{align}
\ln({\Vtot}_{,i}) = \ln(Q_t) - \ln(2\varepsilon)\ C_{t,\text{sample},i} \equiv \func(C_{t,\text{sample},i}) \label{eqnPCRsample}
\end{align}
where ${\Vtot}_{,i}$ is the concentration of copies in sample $i$, given its \qpcr-determined $C_{t,\text{sample},i}$ value. Here, $\ln(Q_t)$ and $\ln(2\varepsilon)$ are two parameters to be estimated as part of the MCMC parameter estimation process described later in this section. As different values for these two parameters are sampled in the MCMC process, the total virus concentration data points vary. The variation in the conversion is denoted by error bars on each total virus data point in Fig.~\ref{fig:fits}~E--G.

\subsection*{Mock-yield assay model}

Loss of virus infectivity or loss of viral genome integrity over time typically follows an exponential decay \cite{beauchemin19}. As such, the mock-yield (MY) or infectivity decay assay can be captured via $V(t) = V_0\, \me{-c\, t}$, such that the experimental MY data are expected to follow
\begin{align}
\ln(\Vinf(t)) &= \ln(V^\text{MY}_{\text{inf},0}) - \cinf\ t \label{eqnMYinf} \\
\ln(\Vtot(t)) &= \ln(V^\text{MY}_{\text{tot},0}) - \ctot\ t \label{eqnMYtot}
\end{align}
where $\Vinf(t)$ and $\Vtot(t)$ are the concentrations of infectious (\tcid/mL) and total (copy/mL) virus after an incubation of time $t$ under the same conditions used during the infection experiments, given the \ebov rate of loss of infectivity ($\cinf$) or integrity ($\ctot$), and initial concentrations, $V^\text{MY}_{\text{inf},0}$ and $V^\text{MY}_{\text{tot},0}$. These data are shown in Fig.~\ref{fig:fits} D, G. 

\subsection*{Simulated infections and parameter estimation}

In estimating the MM parameters, the following experimental data were considered simultaneously: the \qpcr standardized curve (5 data points), the MY assays (24 data points: 4 time points in triplicate for $C_t$ and \Vinf), and three infection assays at MOI of 5 (24 data points: 6 time points in duplicate for $C_t$ and \Vinf), MOI of 1 (54 data points: 9 time points in triplicate for $C_t$ and \Vinf), and MOI of 0.1 (53 data points: 9 time points in triplicate for $C_t$ and \Vinf, minus one contaminated sample in $C_t$).

Eq.~\eqref{eqnPCRstd} was used to capture the \qpcr standard curve, and its agreement with the 5 experimental data points was computed as the sum-of-squared residuals (SSR)
\begin{align*}
\text{SSR}^\text{STD} = \frac{ \sum_{i=1}^5 \left[ \func(C_{t,\text{STD},i}) - \ln(V_{\text{STD},i}) \right]^2 }{\sigma_{\Vtot}^2}\ ,
\end{align*}
where $\sigma_{\Vtot}^2$ is the variance, or squared of the standard error, in experimentally measured $\Vtot$, which will be discussed in more details below. Eq.~\eqref{eqnMYinf} and Eq.~\eqref{eqnMYtot} were used to capture the MY experiment, performed in triplicate, and sampled at 4 time points, for each of $C_t$ and $\Vinf$, and agreement was computed as
\begin{align*}
\text{SSR}^\text{MY} =
\frac{ \sum_{i=1}^{12} \left[ \ln(V^\text{MY}_{\text{inf},0}) - \cinf\ t_i - \ln(\Vinf(t_i)) \right]^2 }{\sigma_{\Vinf}^2} \\
+ \frac{ \sum_{i=1}^{12} \left[ \ln(V^\text{MY}_{\text{tot},0}) - \ctot\ t_i - \func(C_{t,i})\right]^2 }{\sigma_{\Vtot}^2}
\end{align*}
Finally, MM Eq.~\eqref{eqnMM} was used to reproduce the infection experiments, performed in triplicate, and at 3 different MOIs (5, 1, and 0.1), and quantified via both \tcid and \qpcr. In reproducing the infections, initial conditions (at $t=0$) were such that $T(0)=1$, $E_{i=1,...,n_E}(0)=I_{j=1,...,n_I}(0)=0$, and the initial infectious and total virus concentrations for the 3 MOIs were computed as:
\begin{align*}
\Vinf(0) &= V^\text{INF}_{\text{inf},0} \times \text{MOI} \\
\Vtot(0) &= V^\text{INF}_{\text{tot},0} \times \text{MOI}
\end{align*}
where MOI was either 5, 1 or 0.1, and ($V^\text{INF}_{\text{inf},0}$, $V^\text{INF}_{\text{tot},0}$) are 2 parameters to be estimated. Agreement between MM Eq.~\eqref{eqnMM} and experimental infection data was computed as
\begin{align*}
\text{SSR}^\text{INF} = \sum_{\text{MOI}=[5,1,0.1]}
\frac{ \sum_{i} \left[ \ln(\Vinf^\text{MM}(t_i) - \ln(\Vinf(t_i)) \right]^2 }{\sigma_{\Vinf}^2} \\
+ \frac{ \sum_{i} \left[ \ln(\Vtot^\text{MM}(t_i) - \func(C_{t,i}) \right]^2 }{\sigma_{\Vtot}^2}
\end{align*}
where $\sigma_{\Vinf}^2=0.1$ and $\sigma_{\Vtot}^2=0.1$ correspond to the variance in $\ln(\Vinf)$ and $\ln(\Vtot)$, respectively, estimated as the variance of the residuals between the 2 to 3 replicates of $\ln(\Vinf)$ or $\ln(\Vtot)$ measured at each time point and their corresponding mean, across all (STD, MY, and INF) experimental data collected.

A total of 15 parameters --- 6 parameters associated with experimental conditions ($\ln(Q_t)$, $\ln(2\varepsilon)$, $\ln(V^\text{MY}_{\text{inf},0})$, $\ln(V^\text{MY}_{\text{tot},0})$, $V^\text{INF}_{\text{inf},0}$, $V^\text{INF}_{\text{tot},0}$) and 9 parameters more closely associated with \ebov infection kinetics ($\cinf$, $\ctot$, $\pinf$, $\ptot$, $\beta$, $\tau_E$, $\tau_I$, $n_E$, $n_I$) --- were estimated (Table~\ref{table:params}) from 160 experimental data points using the python MCMC implementation phymcmc \cite{phymcmc}, a wrapping library for emcee \cite{emcee}. Posterior probability likelihood distributions (PostPLDs) were obtained based on the parameter likelihood function
\begin{align*}
\ln( \mathcal{L}(\vec{p}) ) = -\frac{1}{2} \left[ \text{SSR}^\text{STD}(\vec{p}) + \text{SSR}^\text{MY}(\vec{p}) + \text{SSR}^\text{INF}(\vec{p}) \right]
\end{align*}
and the assumption of linearly uniform or $\ln$-uniform priors, where $\vec{p}$ is the 15-parameter vector. 

\begin{table}[h!]
\caption{\textbf{Estimated parameters of \ebov infection \textit{in vitro}.} %
}
\label{table:params}
\begin{tabular}{lc} \hline
Parameter & Mode [95\% CR] \\ \hline
Infectiousness, $\beta$ ($\frac{\mathrm{mL}}{\tcid \cdot \mathrm{h}}$) & $10^{-6.48\ [-6.7,-6.3]}$ \\
Eclipse phase length, $\tau_E$ (h) & $30.5\ [26,37]$ \\
Number of eclipse compartments, $n_E$ & $13\ [8,23]$ \\
Infectious phase length, $\tau_I$ (h) & $83.2\ [64,95]$ \\
Number of infectious compartments, $n_I$ & $14\ [3,85]$ \\
Infectious virus production rate, \pinf ($\frac{\tcid}{\mathrm{cell}\cdot\mathrm{h}}$) & $10^{1.12\ [1,1.3]}$ \\
Total virus production rate, \ptot ($\frac{\mathrm{RNA}}{\mathrm{cell}\cdot\mathrm{h}}$) & $10^{6.46\ [6.3,6.7]}$ \\
Rate of loss of infectious virus, \cinf (/h) & $0.0614\ [0.055,0.068]$ \\ 
Rate of virus degradation, \ctot (/h) & $0.00817\ [0.0035,0.013]$ \\ \hline 
Initial infectious virus inoculum, $V^\text{INF}_{\text{inf},0}$ ($\frac{\tcid}{\mathrm{mL}}$) & $10^{5.39\ [5.3,5.5]}$ \\
Initial total virus inoculum, $V^\text{INF}_{\text{tot},0}$ ($\frac{\mathrm{RNA}}{\mathrm{mL}}$) & $10^{11.5\ [11,12]}$ \\ 
MY initial infectious virus inoculum, $\ln(V^\text{MY}_{\text{inf},0})$ & $13.7\ [13,14]$ \\
MY initial total virus inoculum, $\ln(V^\text{MY}_{\text{tot},0})$ & $28.2\ [28,29]$ \\
Standard \qpcr curve $y$-intercept, $\ln(Q_t)$ & $37.8\ [37,39]$ \\
Standard \qpcr curve slope, $\ln(2\varepsilon)$ & $0.613\ [0.57,0.66]$ \\ \hline \hline
Basic reproductive number, $R_0$ & $10^{2.77\ [2.6,3]}$ \\
Infectious burst size, $p_\mathrm{inf}\tau_I$ ($\frac{\tcid}{\mathrm{cell}}$) & $10^{3.04\ [3,3.1]}$ \\
Infecting time, $t_\mathrm{inf}$ (h) & $10^{0.335\ [0.21,0.43]}$ \\ \hline 
\end{tabular}
\end{table}

\section*{Supporting information}
\paragraph*{S1 Appendix.}
\label{S1_Appendix}
{\bf Kinetics of cell-associated virus and cell viability.}

\section*{Acknowledgments}
Members of the CL4 Virology Team include Lin Eastaugh, Lyn M. O’Brien, James S. Findlay, Mark S. Lever, Amanda Phelps, Sarah Durley-White, Jackie Steward and Ruth Thom. The authors would like to acknowledge Joseph Gillard for helpful discussions and the International Centre for Mathematical Sciences (ICMS), where the mathematical model was developed during a Research-in-Groups programme.


\end{document}


\maketitle

\begin{figure}[H]
\centering
\resizebox{\columnwidth}{!}{\includegraphics{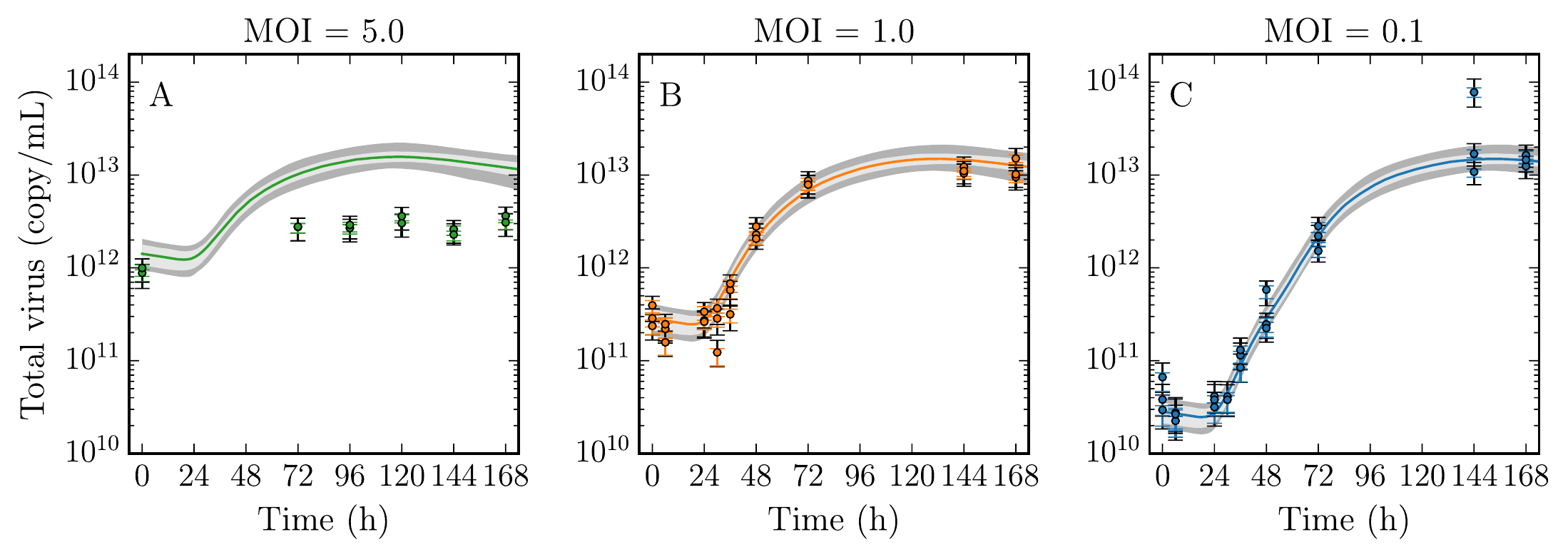}}
\caption{\textbf{Total virus from MOI 5 infection.} %
The total virus concentration from the MOI 5 infection (A) was omitted from the analysis due to an order of magnitude difference in the peak concentration when compared to the MOI 1.0 and 0.1 infections (B, C). It was unclear whether this discrepancy was biologically meaningful or due to systematic experimental error. %
}
\label{fig:moi5RNA}
\end{figure}

\begin{figure}[H]
\centering
\resizebox{\columnwidth}{!}{\includegraphics{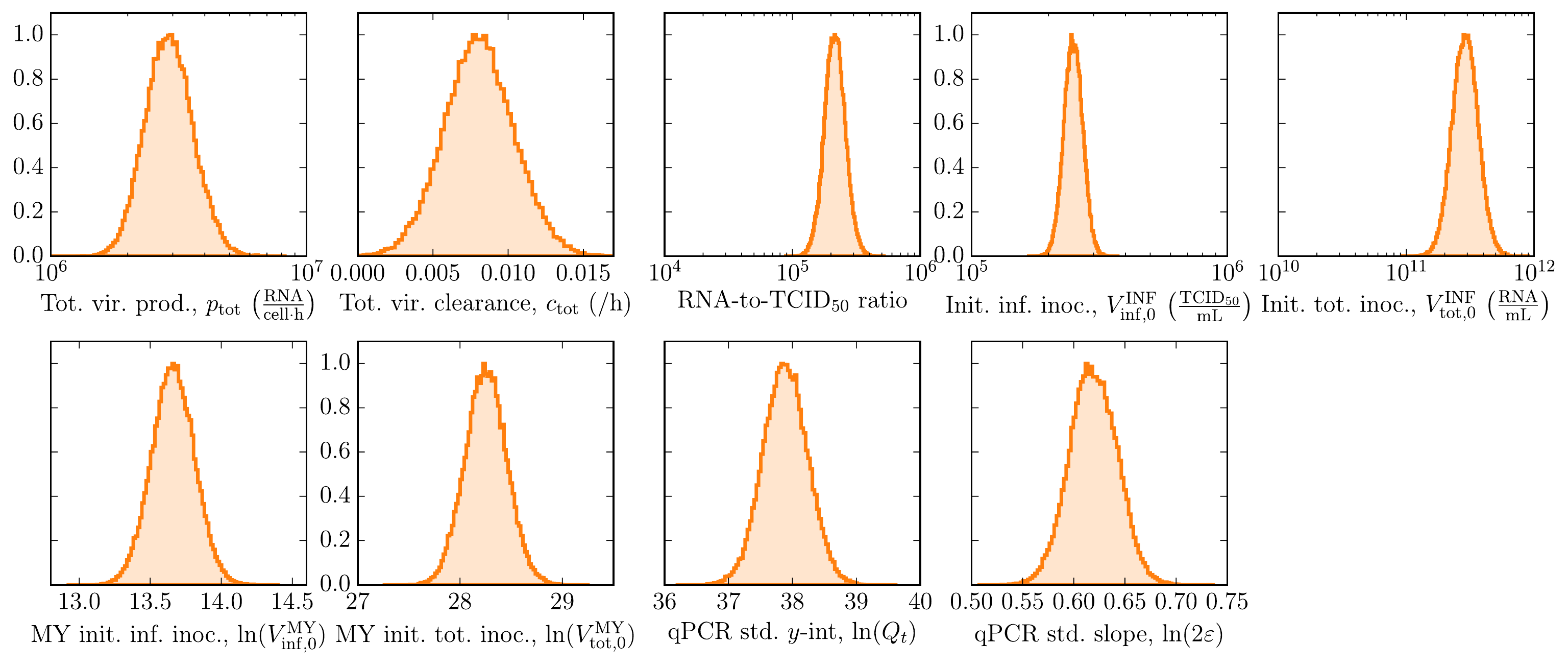}}
\caption{\textbf{Estimated parameter distributions of EBOV infection in vitro.} %
Posterior probability likelihood distributions describing the total virus, mock yield, and calibration curves. %
}
\label{fig:PLDsall}
\end{figure}

\begin{figure}[H]
\centering
\resizebox{\columnwidth}{!}{\includegraphics{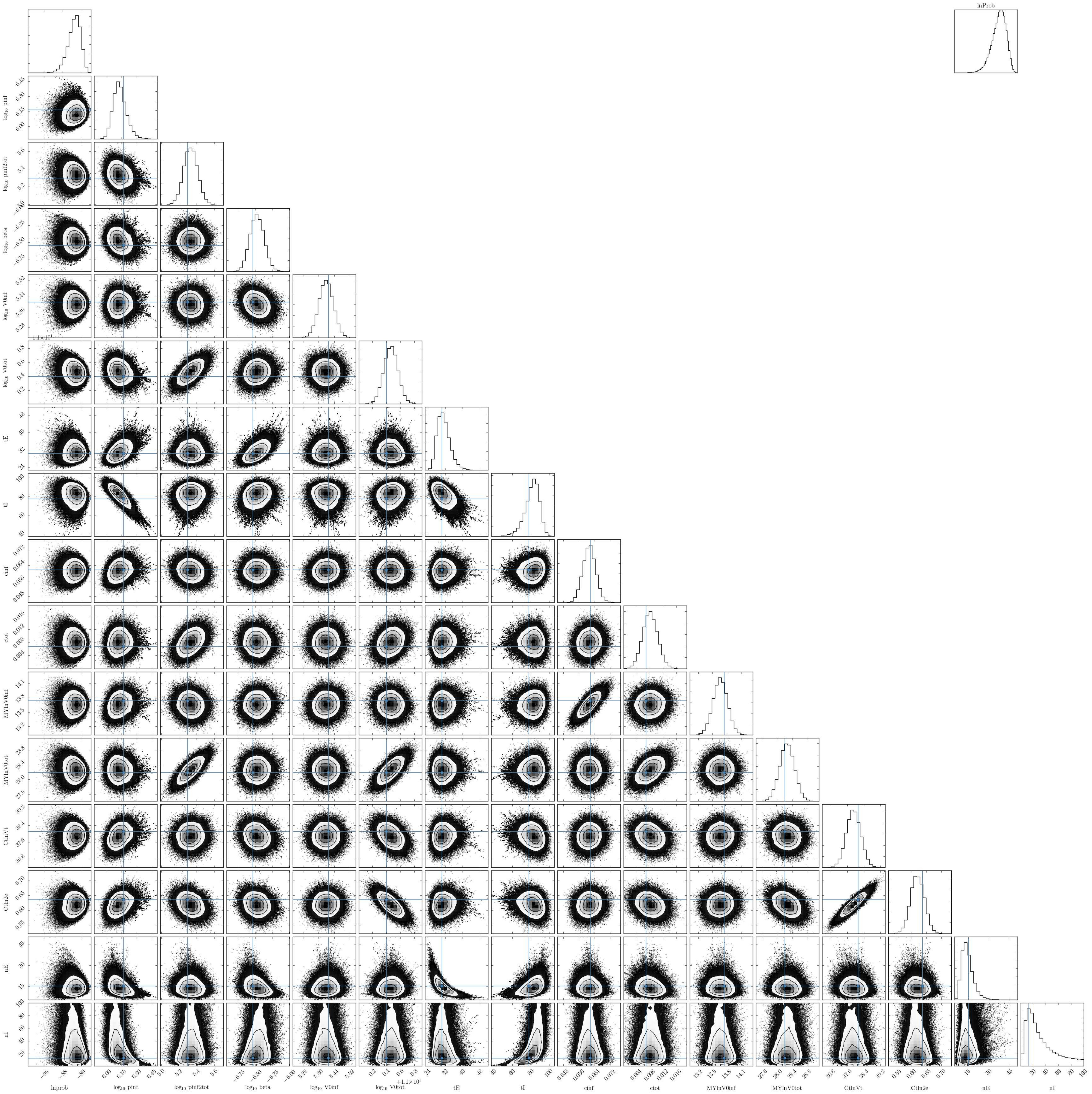}}
\caption{\textbf{Paired parameter posterior likelihood distributions.} %
Two-parameter PLDs for each of the 600,000 MCMC-accepted parameter sets are shown. Although mild correlations were observed, they were not important given that narrow PLDs were obtained. The majority of correlations were a consequence of performing linear regression to the mock yield and standard qRT-PCR curve, resulting in correlations between the (slope, y-int) parameter pairs ($c_\mathrm{inf}$, $\ln(V_\mathrm{inf,0}^\mathrm{MY}$) and ($\ln(2\varepsilon)$, $\ln(Q_t)$), respectively. Since different linear fits to the standard curve impacts the conversion of Ct to RNA, correlations between the total virus parameters are also observed as a result, e.g., ($\ln(2\varepsilon)$, $V_{\mathrm{tot},0}^\mathrm{INF}$), ($\ln(V_{\mathrm{tot},0}^\mathrm{MY})$, $p_\mathrm{tot}$), ($\ln(V_{\mathrm{tot},0}^\mathrm{MY})$, $V_{\mathrm{tot},0}^\mathrm{INF}$). Other correlations are observed such as ($p_\mathrm{inf}$, $\tau_I$), ($n_E$, $\tau_E$), ($n_I$, $t_I$). %
}
\label{fig:triangle}
\end{figure}

\newpage

\section{Kinetics of Cell-associated Virus}

\begin{figure}[H]
\centering
\resizebox{\columnwidth}{!}{\includegraphics{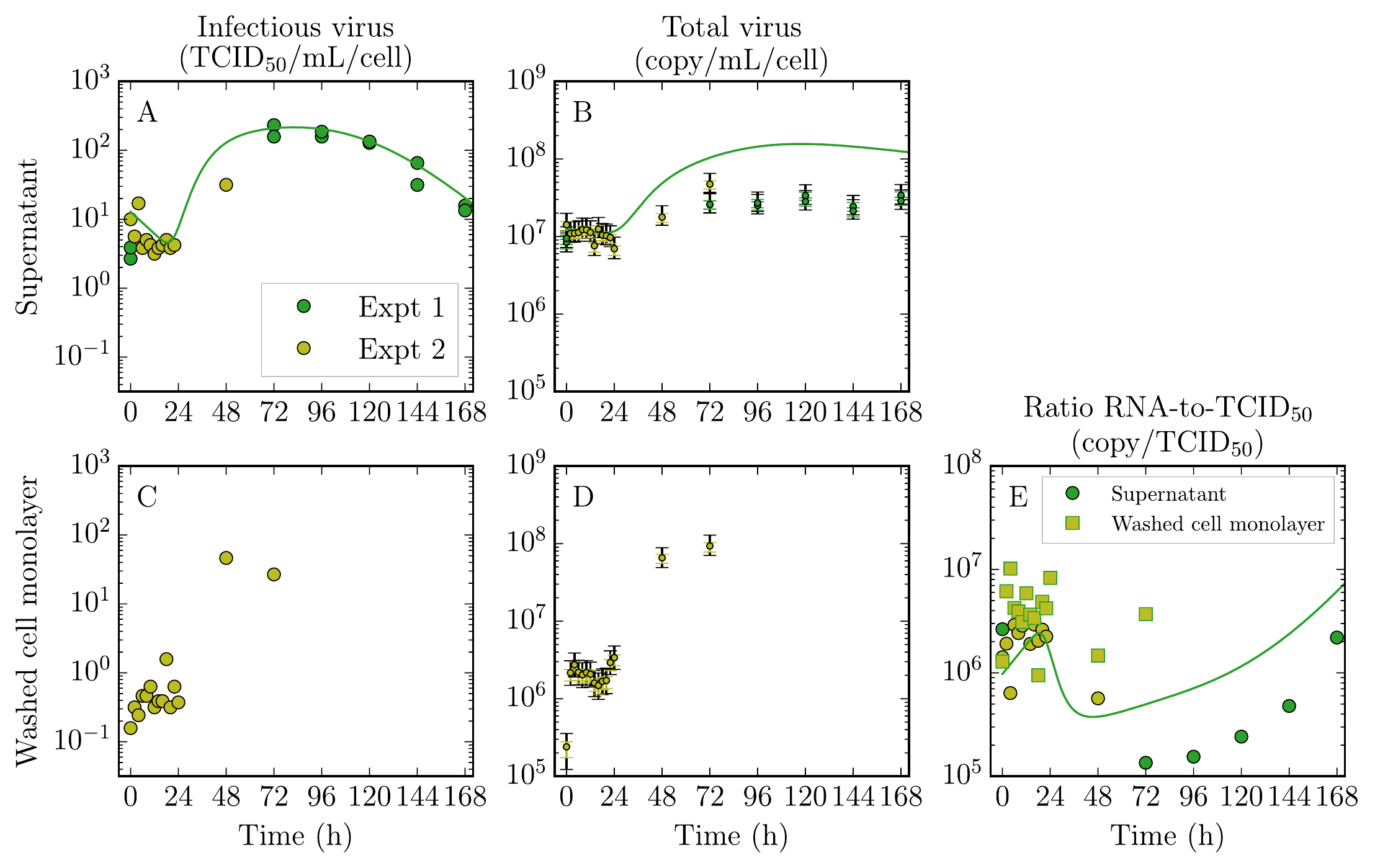}}
\caption{\textbf{Kinetics of MOI 5 infection.} %
In two MOI 5 infection experiments, the infectious and total virus concentrations in the supernatant (A, B) of the in vitro assay were quantified at primarily late times (Expt 1) or early times (Expt 2). In addition, Expt 2 quantified the virus concentrations from washed and trypsinized cell monolayers that remained after removal of the supernatant (C, D). The ratio of RNA-to-\tcid in the supernatant is compared to that in the washed cell monolayers in E. Note that these data have been normalized to the number of cells per well ($10^5$ cells). The lines represent the simulated time course from our MM that corresponds to the set of parameters with the maximum log-likelihood. The variability from converting Ct values to total virus (copy/mL) is shown by two error bars on each total virus data point, denoting the 68\% (same colour) and 95\% (black) CR. %
}
\label{fig:ratios}
\end{figure}

Two MOI 5 infection experiments were performed with cells grown to 90\% and 70\% confluence in Expt 1 and Expt 2, respectively. In Expt 1 (Figure~\ref{fig:ratios}, panels A, B; green), infectious and total virus concentrations were determined from the supernatant of the in vitro assay at times late in the infection (0 and 72--168 hours post-infection). In Expt 2, earlier and much more frequent sampling (every 2 hours from 0--22 hours post-infection) was performed to determine the infectious and total virus concentrations in the supernatant (Figure~\ref{fig:ratios}, panels A, B; yellow). After supernatant removal, additional samples were taken where cell monolayers were washed with PBS, trypsinized and scraped with a pipette tip for total and infectious virus quantification (Figure~\ref{fig:ratios}, panels C, D). 

We expected that the virus concentrations from the washed cell monolayers would correspond to intracellular virus concentrations (Figure~\ref{fig:ratios}, panels C, D). We speculate that the intracellular virus signal was instead obscured by cell-associated virus for the following reasons. Firstly, a very high level of total virus was detected even as early as 0--2 hours post-infection, before significant viral replication was expected to occur. This was accompanied by a very high level of infectious virus which, if we were to believe were due to intracellular virions, is counter to the fact that virions uncoat and release viral RNA once they enter the cell and would not register as infectious virions in a \tcid assay. Lastly, in Figure~\ref{fig:ratios} (panel E), we compared the ratio of RNA-to-\tcid in the supernatant (circles) to that from the washed cell monolayers (squares). The ratio at early times from the washed cell monolayers resembled the ratio in the supernatant. If the washed cell monolayers samples were to reflect the intracellular virus levels, we would expect this ratio to be much higher (more RNA than \tcid within a cell). A likely possibility is that these data reflect the kinetics of cell-associated virions, which are virions that remained attached to the cell membrane even after washing.  

Under the interpretation that these data reflect cell-associated virus kinetics, the high levels of virus at early times are likely virions that adsorbed to the cell membrane. A small jump in virus concentration between 0--2 hours post-infection can be observed, perhaps showing that adsorption is still occurring, until it reaches a steady state by 4 hours post-infection. Once the infected cell comes out of the eclipse phase, by 48 hours post-infection, the ratio of RNA-to-\tcid rises. These data at late times (48, 72 hours post-infection) may no longer be dominated by the cell-associated virus signal, and could contain information on the level of intracellular virus. We did not pursue further mathematical modelling due to the lack of intracellular virus data, beyond these two data points.

Our final analysis only included the infectious virus concentration from Expt 1. There was insufficient information to justify whether the data from Expt 1 and 2 should be combined. We could not verify whether there was consistency in the peak virus concentration, the timing of the rise, or the difference between peak and early virus levels since Expt 2 lacked frequent sampling late enough in the infection to capture such features. As previously mentioned, the total virus concentration from Expt 1 was also omitted. Ultimately, we found that omission of these data did not hamper our ability to robustly extract the viral infection kinetics parameters. 

\newpage

\section{Kinetics of Cell Viability and Infection}

\subsection{Methods}

Following the removal of the supernatant for virus quantification, three wells (two for the ${\rm MOI}=5$ experiment) at each time point were stained with 5\% Trypan blue for $\ge 2$~min, then visualised and photographed at fixed magnification using a Leica DMIRB microscope and the Leica Application Suite v4.9 software.

\begin{figure}[H]
\centering
\resizebox{0.49\columnwidth}{!}{\includegraphics{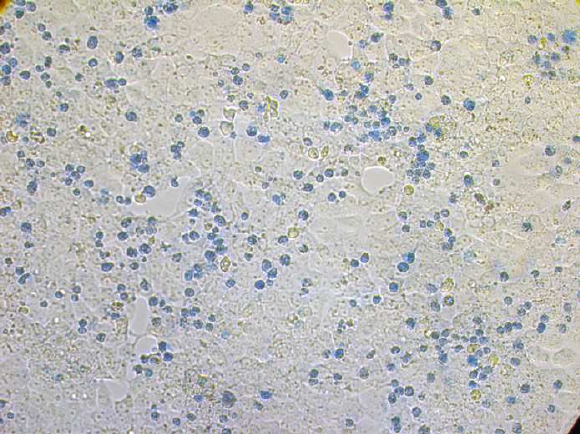}}
\resizebox{0.49\columnwidth}{!}{\includegraphics{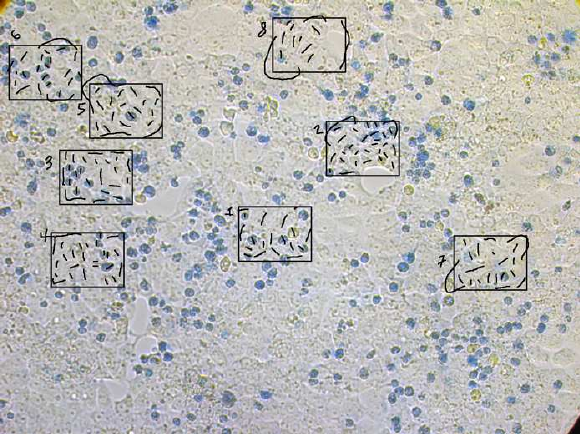}}
\caption{Representative image --- one of the two ${\rm MOI}=5$ images at \unit{168}{h} --- of Trypan blue stained cells in infected monolayers (left), and the counting method used to estimate total number of cells per image (right). Random windows chosen for cell counting were rejected and re-selected if the cells boundaries were not clearly visible. Trypan--blue-stained cells were counted directly.  %
}
\label{fig:cell-count} 
\end{figure}

The total number of stained cells in each image was counted directly since these were generally unambiguously identifiable. The total number of cells per image was estimated by counting intact cells within 5--10 randomly placed windows, $1/64^{\rm th}$ the area of the image, and computing an average (Figure~\ref{fig:cell-count}). Windows in which cell boundaries could not be reliably identified, due to image artifacts or optical effects, were rejected. The mean and standard deviation of each quantity over the replicates was used for analysis.

\subsection{Analysis}

\begin{figure}[H]
\centering
\includegraphics[width=0.9\textwidth]{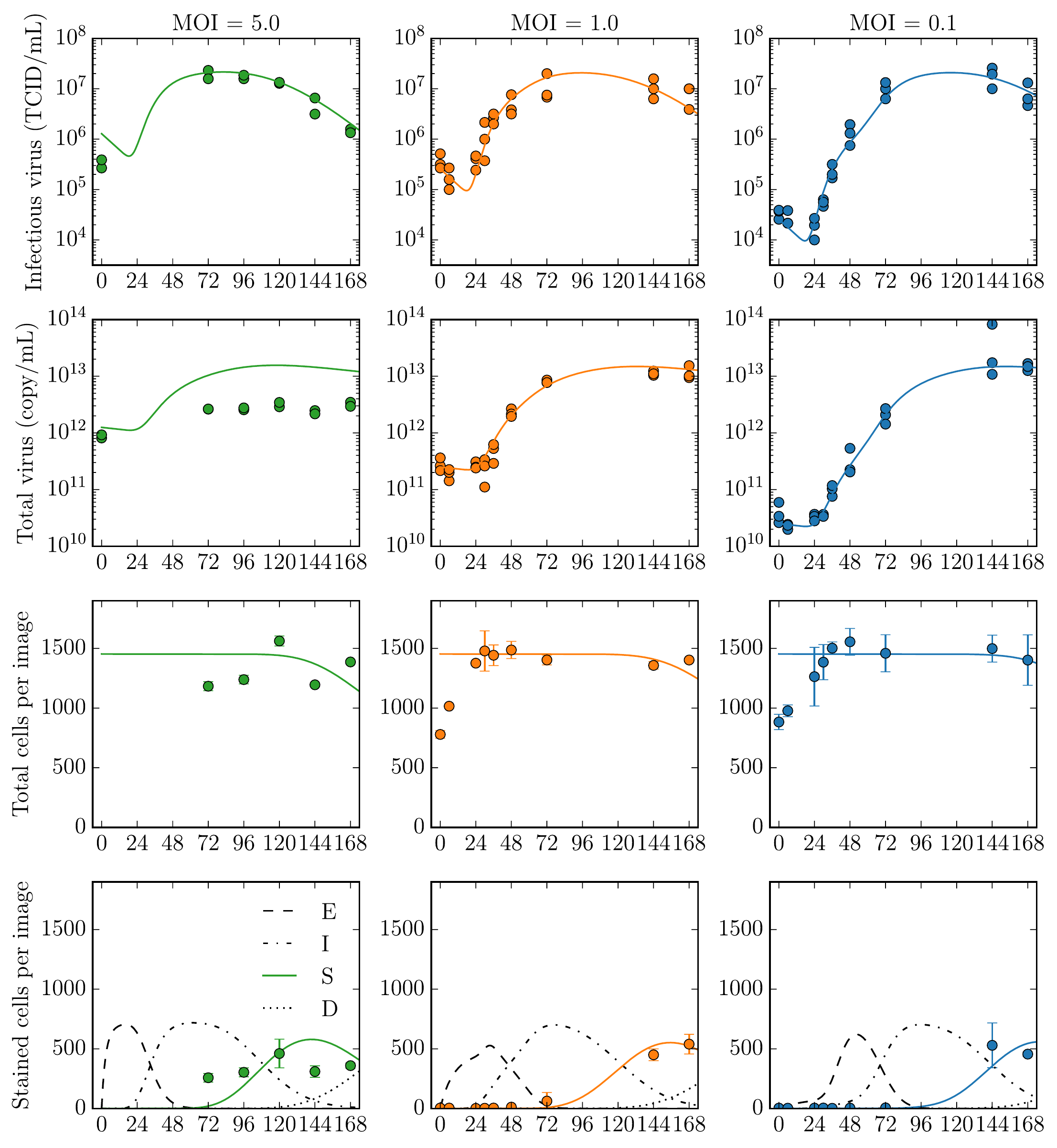}
\caption{Simultaneous fit of viral kinetics data and cell viability kinetics with extended model (see text). To simulate the infection dynamics of the number of cells in each image, we started the infection with $N_\mathrm{image} = 1450$ target cells in the extended model, and rescaled the viral production rates (e.g., $p_\mathrm{inf}\frac{N_\mathrm{well}}{N_\mathrm{image}}$) in order to model the virus concentrations in the entire well. The total cells per image in the model was given by $T+\sum\limits_1^{n_E} E_i + \sum\limits_1^{n_I} I_j + S$. %
}
\label{fig:cell-kinetics} 
\end{figure}

The total cell counts per image revealed a very stable population of intact cells (approximately 1450 cells per image for all three MOI experiments), following a brief period of proliferation in the first day post-infection (Figure~\ref{fig:cell-kinetics}, Total cells per image). In the ${\rm MOI}=1$ experiment, the percent of cells that were stained --- Trypan blue marks cells that have lost the ability to actively exclude the dye --- was low ($\sim5\%$) at \unit{72}{h} post-infection but rose to approximately 30\% by one week post-infection. In the high MOI experiment, 20--30\% of cells were stained by the first measured time point (\unit{72}{h}) and that level was maintained throughout the experiment, while cells infected at low MOI remained unstained at \unit{72}{h}, but were stained at the same 20--30\% level at the next measured timepoint, three days later. 

Despite a nearly constant population size over time, many cell cultures imaged at later times had large holes in the otherwise regular monolayer. These seemingly contradictory observations suggest that, near the end of the one week experiment, a small number of cells disintegrated completely and the monolayer surface tension created larger holes in their absence.

\begin{figure}[H]
\centering
\resizebox{0.5\columnwidth}{!}{\includegraphics{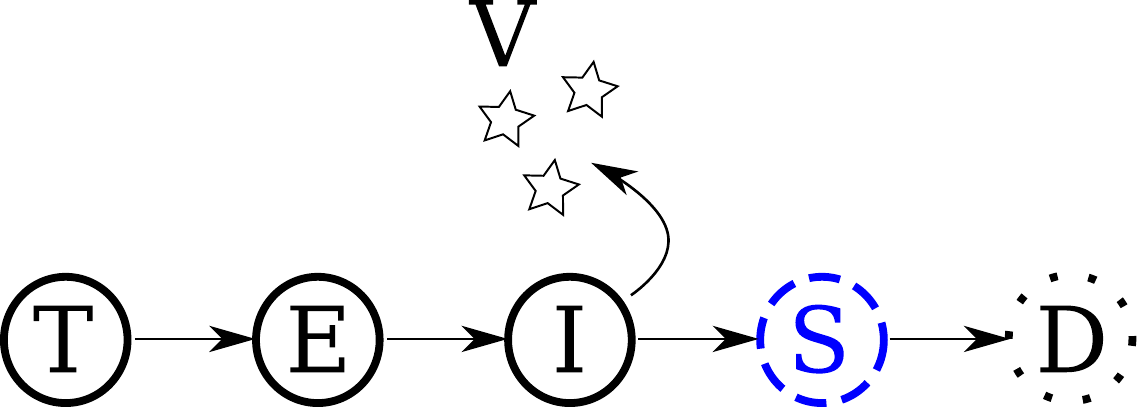}}
\caption{Extended Ebola infection model in which Trypan-blue stained cells ($S$) represent an early phase of cell death prior to their disintegration into uncountable debris ($D$). %
}
\label{fig:model1} 
\end{figure}
To simulate these cell viability kinetics in tandem with our infection model (presented in the main text) we assumed that once infected cells cease viral production, they enter a relatively long-lasting phase where their membrane is permeable to Trypan blue and they become stained (S) cells (Figure~\ref{fig:model1}). As the stages of cell death progress, stained cells subsequently disintegrate to debris (D) and cannot be counted. Representing the stained phase with a sequence of $n_{\rm S}$ equations to obtain Erlang-distributed timings, i.e.,
	\begin{equation*}
	\begin{split}
	\frac{{\rm d}S_1}{{\rm d}t} &= \frac{n_I}{\tau_I} I_{n_I} - \frac{n_S}{\tau_S} S_1\\
	\frac{{\rm d}S_{j=2,3,\ldots,n_S}}{{\rm d}t} &= \frac{n_S}{\tau_S} \left( S_{j-1} - S_j \right)\,,
	\end{split} 
	\end{equation*}
we found that an average S time of \unit{70}{h} ($n_{\rm S} = 10$) was sufficient to reproduce the observed constant total cell count per image over the one week experiment (Figure~\ref{fig:cell-kinetics}, Total cells per image). Finally, to account for the fact that stained cells made up only a modest fraction of the total number of cells per image, even at late times, we allowed for only 50\% of the target cells to be susceptible to infection. Under these assumptions, we obtained an adequate fit to the stained cell data for all three MOI experiments (Figure~\ref{fig:cell-kinetics}, Stained cells per image) while holding all but one of the infection parameters (the viral production rate was doubled to account for the halved susceptible population) at their maximum likelihood values (Table 1, main text), thus maintaining agreement with the viral kinetics data.

\subsection{Discussion}

In our final analysis (main text), we chose to exclude cell data from consideration because, at least in its simplest interpretation, it had no effect on the estimation of the viral kinetics parameters (the only exception being the viral production rate, which could differ by a factor of two, as discussed above). Moreover, any additional information that could be determined about the viability kinetics of EBOV-infected cells would be based solely on assumptions about how the timing of Trypan blue staining fits within the infection timeline. The Trypan blue dye stains cells that can no longer actively exclude it, implying that stained cells are dead. But the onset of this passive permeability with respect to cessation of viral production (the end of the ``infectious'' phase of infection) is not known, and therefore acts as a hidden parameter (see below). Future resolution of this problem, or the use of a staining method that can be registered to the infection timeline, e.g., immuno-staining, could potentially allow for these cell kinetics data to further constrain the viral kinetics parameters.

\begin{figure}[H]
\centering
\resizebox{0.6\columnwidth}{!}{\includegraphics{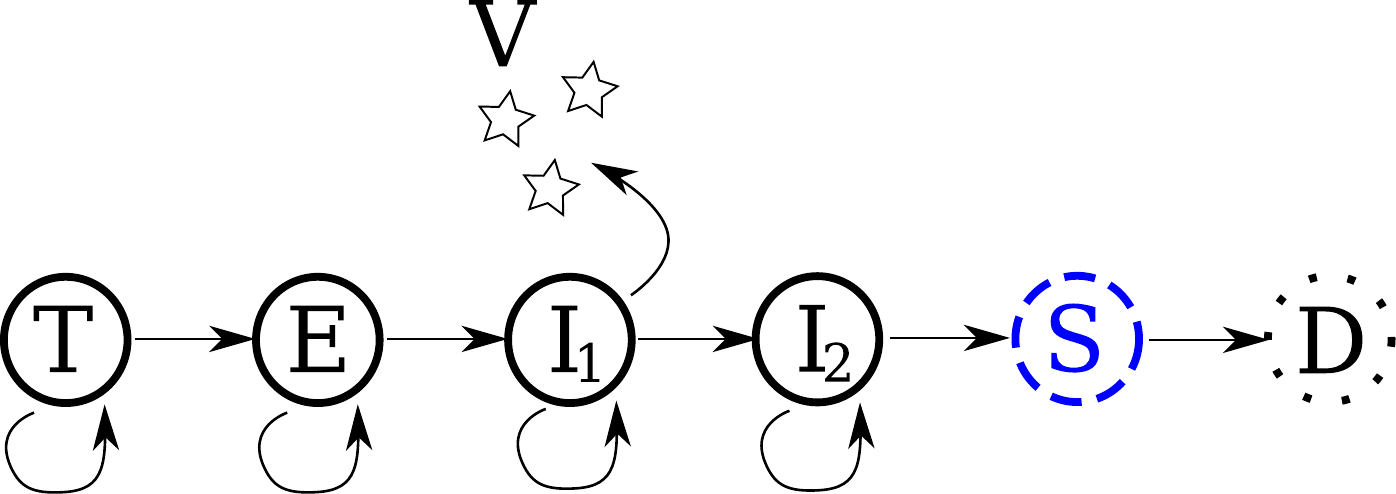}}
\caption{Alternative model in which a proliferating ``post-infectious'' phase ($I_2$) continuously feeds a quasi--steady-state population of stained cells ($S$). Tuning the relative lifespans of each population can yield the observed fractional level of the stained population, without the requirement that some target cells are not susceptible to infection. %
}
\label{fig:model2} 
\end{figure}

We were able to obtain slightly better agreement to the cell kinetics data (not shown) with additional model features.  Adding logistic proliferation for target and eclipse cells to the above-described model (with a growth rate $r=\unit{0.105}{{\rm h}^{-1}}$ common to all three MOI experiments)  allowed for a very good fit to the first \unit{24}{h} of cell data, without altering the agreement to the stained cell or viral kinetic data. To avoid the assumption that only half of the target cells are susceptible, an alternative model (Figure~\ref{fig:model2}) could be used where one assumes:  (i) a post-infectious phase ($I_2$) in which infected cells no longer produce virus but continue to actively exclude dye, and (ii) that all living cells, including those in the post-infectious phase, proliferate. This results in a model with the following set of equations for the post-infectious phase:
	\begin{equation*}
	\begin{split}
	\frac{{\rm d}I_{21}}{{\rm d}t} &= \frac{n_{I_1}}{\tau_{I_1}} I_{1n_{I_1}} - \frac{n_{I_2}}{\tau_{I_2}} I_{21} + r\, I_{21} \left(1 - \frac{T+\sum E_j + \sum I_{1j} + \sum I_{2j}}{N_{\rm max}}\right) \\
	\frac{{\rm d}I_{2j=2,3,\ldots,n_{I_2}}}{{\rm d}t} &= \frac{n_{I_2}}{\tau_{I_2}} \left(I_{2j-1} - I_{2j} \right) + r\, I_{2j} \left(1 - \frac{T+\sum E_j + \sum I_{1j} + \sum I_{2j}}{N_{\rm max}}\right) \,,
	\end{split} 
	\end{equation*}
and other model equations adapted accordingly. Under these assumptions the stained (S) cells can form a late-time quasi--steady-state population that is constantly replenished by ongoing proliferation, whose fractional level with respect to the total population can be tuned by the relative rates of entry and exit to/from the S phase. This allows for a better fit to the stained cell data than that shown in Figure~\ref{fig:cell-kinetics} (Stained cells per image), but requires a re-tuning of infection parameters to maintain agreement with the viral kinetics data. Given the current lack of knowledge about the end stage viability of infected cells, and the ambiguity of the Trypan blue stain, described above, we did not pursue these more complicated models any further.